\documentclass[10pt,letterpaper]{article}
\usepackage{opex3}

\begin{document}

\title{Laser written waveguide photonic quantum circuits}

\author{Graham~D.~Marshall,$^{1,\dagger,\star}$ Alberto~Politi,$^{2,\star}$ Jonathan~C.~F.~Matthews,$^{2,\star}$ Peter~Dekker,$^1$ Martin~Ams,$^1$ Michael~J.~Withford,$^1$ and Jeremy~L.~O'Brien$^{2,\ddagger}$ }
\address{$^1$Centre for Ultrahigh bandwidth Devices for Optical Systems (CUDOS), MQ Photonics Research Centre, Department of Physics, Macquarie University, NSW 2109, Australia}
\address{$^2$Centre for Quantum Photonics, H. H. Wills Physics Laboratory \& Department of Electrical and Electronic Engineering, University of Bristol, Merchant Venturers Building, Woodland Road, Bristol, BS8 1UB, UK}
\address{$^\star$These authors contributed equally to this work}


\begin{center}
        \begin{minipage}[t]{0.4\textwidth}
            \email{$^\dagger$graham@science.mq.edu.au}
            \vspace{-12pt}
            \homepage{$^\dagger$www.ics.mq.edu/cudos}
        \end{minipage}
        \begin{minipage}[t]{0.4\textwidth}
            \email{$^\ddagger$jeremy.obrien@bristol.ac.uk}
            \vspace{-12pt}
            \homepage{$^\ddagger$www.phy.bris.ac.uk/groups/cqp}
        \end{minipage}
\end{center}

\begin{abstract}
We report photonic quantum circuits created using an ultrafast laser processing technique that is rapid, requires no lithographic mask and can be used to create three-dimensional networks of waveguide devices. We have characterized directional couplers---the key functional elements of photonic quantum circuits---and found that they {perform as well as} lithographically produced waveguide devices. We further demonstrate high-performance interferometers and an important multi-photon quantum interference phenomenon for the first time in integrated optics. This direct-write approach will enable the rapid development of sophisticated quantum optical circuits and their scaling into three-dimensions.
\end{abstract}






\section{Introduction}
\vspace{6pt}
Quantum information science promises exponential improvement and new functionality for particular tasks in computation \cite{nielsen}, metrology \cite{gi-sci-306-1330}, lithography \cite{bo-prl-85-2733} and communication \cite{gi-rmp-74-145,gi-nphot-1-165}. Photonics appears destined for a central role owing to the wide compatibility, low-noise and high-speed transmission properties of photons \cite{gi-nphot-1-165,ob-sci-318-1567}. However, future quantum technologies and fundamental science will require integrated optical circuits that offer high-fidelity and stability whilst enabling scalability. Silica-on-silicon waveguide photonic quantum circuits \cite{po-sci-320-646} are an important step, however, conventional lithography is costly, time consuming, and limited to two-dimensions. Here we demonstrate an alternative fabrication technique based on ultrafast laser processing \cite{da-ol-21-1729,no-apa-77-109} that overcomes all of these issues.

Quantum technologies rely on transmitting and processing information encoded in physical systems---typically two-state \emph{qubits}---exhibiting uniquely quantum mechanical properties \cite{nielsen}. Photons hold great promise as qubits given their light-speed transmission, ease of manipulation at the single qubit level, low noise (or \emph{decoherence}) properties and the multiple degrees of freedom available for encoding qubits or higher-level systems (including polarization, optical mode, path and time-bin).

The problem of realizing interactions between single photonic {qubits} was theoretically solved in a breakthrough scheme for implementing optical non-linear interactions using only single photon sources, linear optical elements and single photon detectors \cite{kn-nat-409-46}. Remarkable progress in the development of single photon sources \cite{sh-nphot-1-215} and detectors \cite{ta-nphot-1-343,ga-nphot-1-585}, make a photonic approach to quantum technologies very promising. Indeed, there have been a number of important proof-of-principal demonstrations of quantum optical circuits for communication \cite{pr-prl-94-220406,ya-nphot-2-488}, computing \cite{ob-sci-318-1567}, metrology \cite{mi-nat-429-161,na-sci-316-726,re-prl-98-223601}, and lithography \cite{da-prl-87-013602}. However, the use of large-scale (bulk) optics to create these circuits places extremely stringent requirements on the alignment and stability in position of the optical components, thus making such an approach inherently unscalable. Successful operation of a quantum optical circuit requires that the individual photons be brought together at precisely the same position and with the correct phase on a succession of optical components in order to realize the high-fidelity classical and quantum interference that lies at the heart of single photon interactions \cite{kn-nat-409-46}. Integrated optical quantum circuits based on chip-scale waveguide networks will likely find important applications in future quantum information science along side optical fibre photonic quantum circuits that have been demonstrated in quantum key distribution \cite{gi-nphot-1-165} and quantum logic gate applications \cite{clark-2008}.

Ultrafast lasers are a powerful tool not only for machining \cite{ga-nphot-2-219} but also for the subtle optical modification of materials. In particular, the direct-write femtosecond laser technique for creating optical waveguides in dielectric media \cite{da-ol-21-1729} is an alternative waveguide manufacturing technique that allows the production of low-volume complex three-dimensional optical \emph{circuits} (Fig.~\ref{schematic}(a)). This process has been applied to a wide range of passive and active media to create integrated devices such as microfluidic sensors \cite{os-apl-90-231118}, waveguide-Bragg gratings \cite{ma-ol-31-2690} and miniature lasers \cite{ma-ol-33-956}. Because there is no lithography step in this procedure it enables a waveguide circuit to be taken rapidly from concept to a completed device. However, as with all waveguide fabrication processes, the  devices are subject to manufacturing imperfections and there has been no previous demonstration that the use of the laser-writing technique can produce waveguides that can operate on single photons without deleterious effects on phase, spatial mode or polarization. {In this paper we report the first application of laser written waveguide circuits to photonic quantum technologies and, using single- and multi-photon interference experiments, we show that such circuits perform as well as lithographically fabricated devices and are an ideal platform for scalable quantum information science.}
\vspace{6pt}
\section{Experimental techniques}
\vspace{6pt}
\subsection{Waveguide fabrication}
\vspace{6pt}
In conventional lithographically fabricated integrated optical devices, light is guided in waveguides consisting of a core and a slightly lower refractive-index cladding or buffer layers (in a manner analogous to an optical fibre). In the commonly used flame hydrolysis deposition (FHD) fabrication method these structures are lithographically described on top of a semiconductor wafer \cite{po-sci-320-646}. By careful choice of core and cladding dimensions and refractive index difference it is possible to design such waveguides to support only a single transverse mode for a given wavelength range. Coupling between waveguides, to realize beamsplitter-like operation, can be achieved when two waveguides are brought sufficiently close together that the evanescent fields overlap; this architecture is known as the directional coupler. By carefully selecting the separation between the waveguides and the length of the interaction region the amount of light coupled from one waveguide into the other (the coupling ratio $1-\eta$, where $\eta$ is equivalent to beamsplitter reflectivity) and its dependance on wavelength can be tuned. A similar approach can be taken in the case of directly written waveguides, where the waveguide core is formed by local modification of silica \cite{chan2003a,li-oe-16-20029} (or other materials). However, unlike the lithographic approach, direct-write circuits can be straightforwardly written in 3D.

\begin{figure}[t]
    \begin{center}
        \begin{minipage}[t]{0.02\textwidth}
             \vspace{0pt}
            \hspace{-6pt}
             (a)
        \end{minipage}
        \begin{minipage}[t]{0.3\textwidth}
            \vspace{0pt}
            \hspace{0pt}
            \includegraphics*[width=1.0\textwidth]{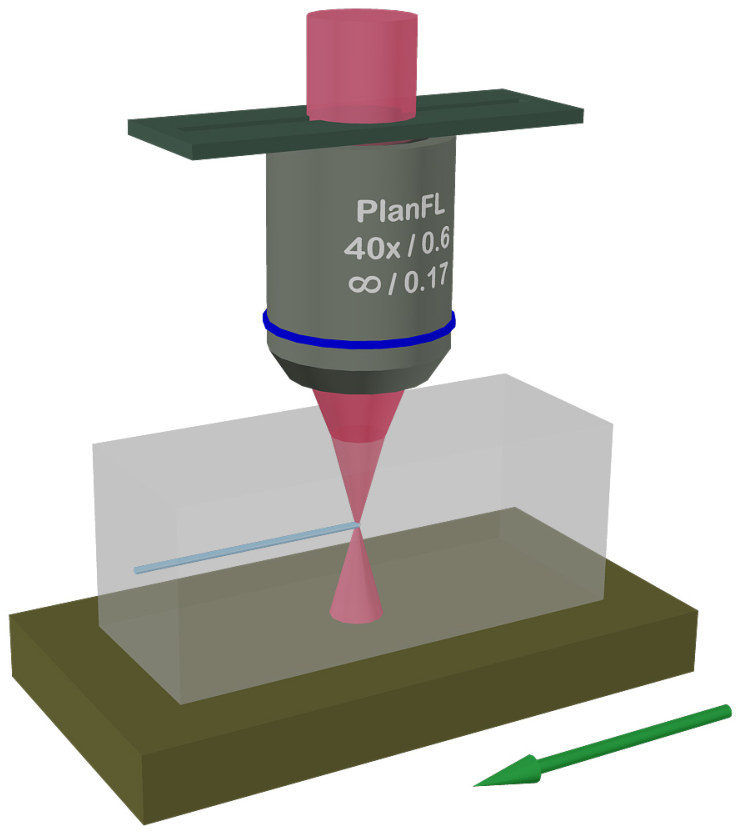}
        \end{minipage}
        \begin{minipage}[t]{0.02\textwidth}
             \vspace{0pt}
            \hspace{-6pt}
             (b)
        \end{minipage}
        \begin{minipage}[t]{0.45\textwidth}
            \vspace{-18pt}
            \hspace{0pt}
            \includegraphics*[width=1.0\textwidth]{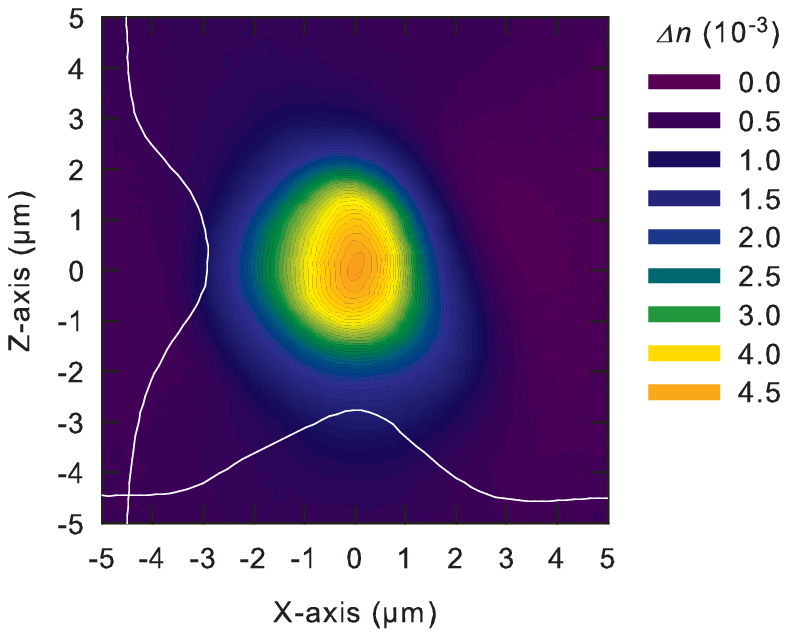}
        \end{minipage}

        \begin{minipage}[t]{0.02\textwidth}
            \vspace{0pt}
            \hspace{-12pt}
            (c)
        \end{minipage}
        \begin{minipage}[t]{0.3\textwidth}
            \vspace{0pt}
            \hspace{0pt}
            \includegraphics*[width=1.0\textwidth]{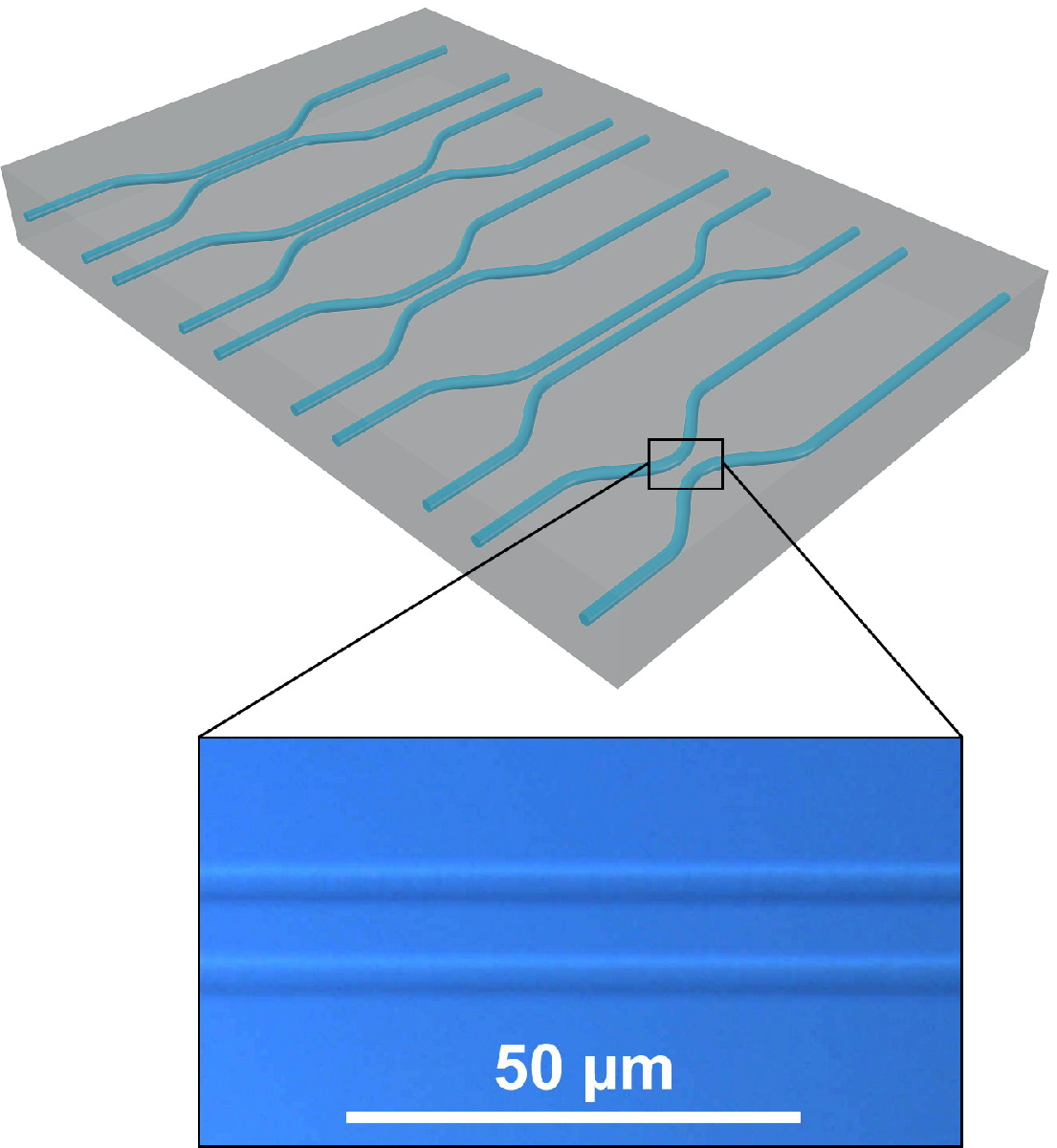}
        \end{minipage}
        \begin{minipage}[t]{0.02\textwidth}
            \vspace{0pt}
            \hspace{-12pt}
            (d)
        \end{minipage}
        \begin{minipage}[t]{0.375\textwidth}
            \vspace{0pt}
            \hspace{0pt}
	       \includegraphics*[width=1.0\textwidth]{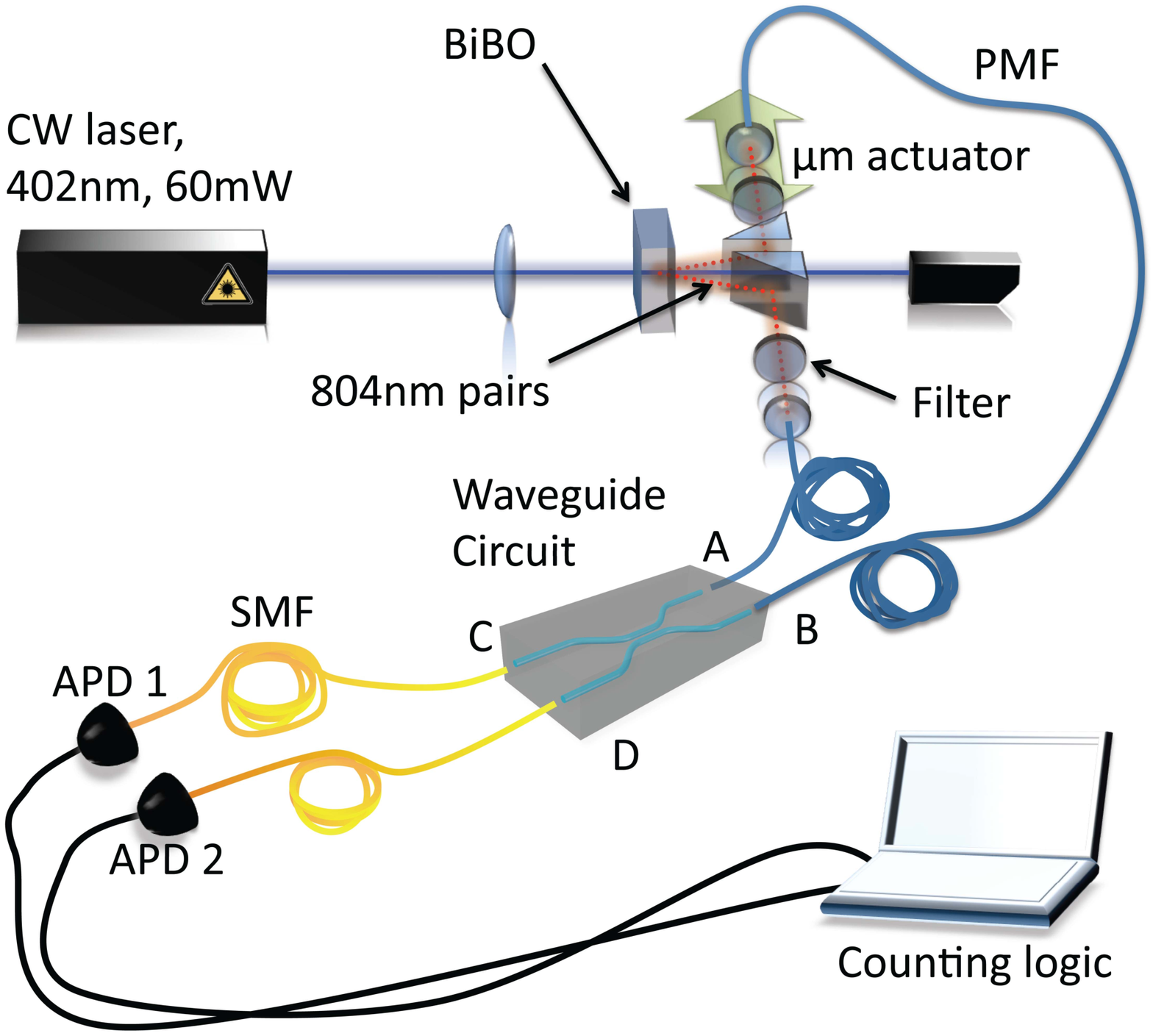}
        \end{minipage}
\caption{Fabrication and measurement of laser direct-write photonic quantum circuits. (a) A schematic of the femtosecond-laser direct-write process. {(b) A refractive index (RI) profile of a typical waveguide where the writing laser was incident from the left. The white overlayed plots show the cross section of the RI profile at the peak. The distortion to the measured $x$-axis RI profile to the right of the peak is an artifact of the measurement method.} (c) A schematic of an array of directional couplers fabricated by \textit{fs} direct writing in a single fused silica chip and an optical micro-graph showing the central coupling region where the waveguides are separated by 10~$\mu$m. (d) Measurement setup based on spontaneous parametric down conversion of a continuous wave 402~nm laser diode. }
\label{schematic}
\end{center}
\end{figure}

We fabricated two chips with a number of direct-write quantum circuits (DWQCs) composed of 2$\times$2 directional couplers (Fig.~\ref{schematic}(c)) and Mach-Zehnder interferometers. The circuits were written inside high purity fused silica using a tightly focused {1~kHz repetition rate, 800~nm, 120~fs laser} and motion control system similar to that previously reported \cite{ams2005,ams2006}. The writing laser beam was circularly polarised and passed through a 520 $\mu$m slit before being focused 170~$\mu$m below the surface of the glass using a 40$\times$ 0.6 numerical aperture microscope objective that was corrected for spherical aberrations at this depth (Fig.~\ref{schematic}(a)). The writing process created approximately {Gaussian profile waveguides which were characterised using a near-field refractive index profilometer (from Rinck Elektronik) and a magnifying beam profiler. The \textit{measured} refractive index profile of a typical waveguide is displayed in Fig.~\ref{schematic}(b) and shows a peak refractive index change of 4.55$\times10^{-3}$, an average $1/e^2$ width of 5.5~$\mu$m and a $x/z$ width ratio of 0.93. At their design wavelength of 806~nm these waveguides supported a single transverse mode with orthogonal (intensity) $1/e^2$ widths of 6.1~$\mu$m~$\times$~6.1~$\mu$m (the small amount of measured RI shape-asymmetry not being evident in the guided mode)}. The design of the directional couplers was functionally identical except for the length of the central {interaction region that was varied from 400 to 2000~$\mu$m} to achieve different coupling ratios. The curved regions of the waveguides were of raised-sine form and connected the input and output waveguide pitch of 250~$\mu$m down to the closely spaced evanescent coupling region of the devices. {The Mach-Zehnder interferometers' design comprised two 50:50 directional couplers separated by identical 1500~$\mu$m long arms. The purpose of these devices was to test the stability of both the completed waveguide circuits and the laser writing system (which was required to remain stable for the several hours required for fabrication of the directional couplers and interferometers).}

\vspace{6pt}
\subsection{Single and 2-photon quantum interference}
\begin{figure}
\begin{center}
\includegraphics*[width=0.45\textwidth]{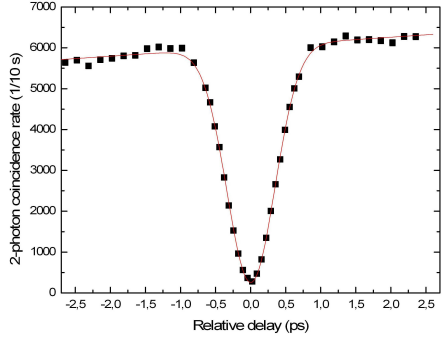}
\caption{Quantum interference in a laser direct-write directional coupler.  The number of coincident detections are shown as a function of the arrival delay between the two interfering photons. Error bars from Poissonian statistics are smaller than the point size and the fit is a Gaussian plus linear.}
\label{dip}
\end{center}
\end{figure}

\vspace{6pt}
The quantum circuits were studied using {photons} obtained using spontaneous parametric down conversion (SPDC) sources. For the two-photon interference work the output from a continuous wave (CW) 402~nm laser diode was down converted into two unentangled 804~nm photons in a type-I phase matched BiBO crystal (Fig.~\ref{schematic}(d)). The photon pairs were passed through 2~nm bandpass interference filters to improve indistinguishability before being coupled into two polarization maintaining single mode optical fibres (PMFs). The path length difference could be varied with the $\mu$m actuator which adjusted the arrival time of the photons with respect to each other at the waveguide chip. Photons were collected from the chip in single mode fibres (SMFs) and coupled to avalanche photo diodes (APDs) which were in turn connected to a photon counting and coincidence logic circuit. Index matching oil was used between the fibres and the device under test to reduce Fresnel reflections that contribute to coupling losses. The circuit devices had typical transmission efficiencies of 50\% (which includes coupling and propagation losses).
In the case of an ideal $\eta=\frac{1}{2}$ directional coupler, if two non-degenerate (\textit{i.e.} entirely distinguishable) photons are coupled into waveguides $A$ and $B$ (Fig.~\ref{schematic}(d)) the photons have a 50\% probability of both being transmitted or reflected inside the coupler---\textit{i.e.}, they behave like ``classical" particles---such that there is a 50\% probability of detecting them simultaneously at the two different APDs. In contrast, if two degenerate photons are input to $A$ and $B$ the pair will be transformed into an entangled superposition of two photons in output waveguide $C$ and two photons in output waveguide $D$:
\begin{equation}
\label{hom}
|11\rangle_{AB}\rightarrow\frac{|20\rangle_{CD}-|02\rangle_{CD}}{\sqrt{2}}.
\end{equation}
This \emph{quantum interference} ideally yields no simultaneous photon detection events at the separate APDs when the difference in arrival time of the photons at the coupler is zero, giving rise to the characteristic Hong-Ou-Mandel \cite{ho-prl-59-2044} (HOM) ``dip" in the coincident detection rate as a function of delay (see Fig.~\ref{dip}). {In our setup, the relative arrival time of the two photons was the free parameter in photon distinguishability and allowed the level of quantum interference to be directly controlled. More generally the visibility of the HOM dip is also limited by other degrees of freedom including polarization, frequency, spatial mode, and beam coupling ratio (or reflectivity) of the directional coupler. While the reflectivity parameter is inherent to the optical circuitry, the remaining degrees of freedom can be attributed to both the source of photons and individual manipulation of these parameters within the circuit.}

{Considering the device separately to the source of photons,} the \textit{ideal} HOM dip visibility, $V\equiv (max-min)/max$, is a function of the equivalent beamsplitter reflectivity $\eta$:
\begin{equation}
\label{visibility}
V_{ideal}=\frac{2\eta(1-\eta)}{1 - 2 \eta + 2 \eta^2}
\end{equation}
{which is a maximum for $\eta=\frac{1}{2}$}. Imperfections in the waveguides {that perturb the state of a photon in any degree of freedom} will degrade the degeneracy of the photon pairs and reduce the measured $V$ below $V_{ideal}$. {Assuming a SPDC source prepares identical photon pairs the relative visibility $V_{rel}\equiv V/V_{ideal}$ of the HOM dip provides the operational fidelity of a directional coupler and thus $V_{rel}$ is} the key quantifier of the performance of a photonic quantum circuit {in preserving photon degeneracy. In this work we have not corrected for imperfections in either the sources' degeneracies or the measurement setups, hence our measurements of $V_{rel}$ are lower bounds for the fidelities of the devices.}

In addition to high-fidelity quantum interference (as quantified by $V_{rel}$), quantum circuits require high-visibility, stable classical interference. {Using the 2-photon SPDC source as a convenient source of 804~nm photons we measured the effective reflectivity of the Mach-Zehnder interferometers by blocking one arm of the SPDC source to input one photon at a time into the arms of the interferometer. In the case of an ideal null-phase difference interferometer with 50:50 beam splitters the effective reflectivity of the device, $\eta_{MZ}$, should be 1. Values for $\eta_{MZ}<1$ or instabilities under changing environmental conditions can therefore be used as a device performance metric.}
\vspace{6pt}
\subsection{3-photon quantum interference}
\vspace{6pt}
{In general, quantum circuits for photonic quantum technologies and other applications involve not just one- and two-photon (quantum) interference as described above, but generalized quantum interference of multiphoton inputs to a beamsplitter. In particular quantum interference of multiple photons at a beamsplitter is crucial in applications such as} quantum logic gates \cite{kn-nat-409-46,sa-prl-92-017902}, quantum metrology \cite{li-pra-77-023815}, photon number filters \cite{sa-prl-96-083601,re-prl-98-203602}, entanglement filters \cite{ho-prl-88-147901,ok-sci-323-483}, and biphoton qutrit unitaries \cite{la-prl-100-060504}. When two photons are input into $A$ and one in $B$ an ideal $\eta=2/3$ reflectivity coupler will generate the three-photon entangled state:
\begin{equation}
\label{21hom}
|21\rangle_{AB}\rightarrow\frac{2}{3}|30\rangle_{CD}-\frac{\sqrt{3}}{3}|12\rangle_{CD}-\frac{\sqrt{2}}{3}|03\rangle_{CD},
\end{equation}
where quantum interference results in no $|21\rangle_{CD}$ term. An analogue of the HOM dip can therefore be observed by measuring the rate of detecting two photons in C and one in D as a function of the delay time between the photon in $B$ and the two photons in $A$ \cite{sa-prl-96-083601}. To observe a $|21\rangle_{CD}$ HOM dip, as described by Eq.~(\ref{21hom}), we used a pulsed laser system (Fig.~\ref{21dip}(a)) to generate four photons in 2 modes at 780~nm where the DWQCs are also single moded. The output of a $\sim$150~fs, 80~MHz repetition rate 780~nm Ti:Sapphire laser was frequency doubled to 390~nm and then down converted into pairs of 780~nm photons in a type-I phase matched BiBO crystal. The photon pairs passed through 3~nm bandpass interference filters before being coupled into two polarization maintaining single mode optical fibres. In all other respects the setup is similar to the CW one shown in Fig.~\ref{schematic}(d). By using a fused PMF splitter and single photon APD we were able to probabilistically prepare the $|1\rangle_B$ state at input $B$ (Fig.~\ref{21dip}(a)). Using a SMF fibre coupler we were able to probabilistically detect two photons at output $C$. {The low probability of preparing the required 4-photon state at the source and the $\frac{1}{4}$ success rate of experimental setup significantly reduced the measurement count rate from that of the 2-photon experiment. By testing circuit stability over long durations, interference measurements of this form are a crucial trial of these circuits for applications in advanced, multi-photon quantum optics. Each set of measurements with this setup took $\sim$60~hours to complete and are the first demonstration of interference between more than 2-photons in an integrated waveguide-chip platform.}
\vspace{6pt}
\section{Results}
\begin{figure}
\begin{center}
\includegraphics*[width=0.45\textwidth]{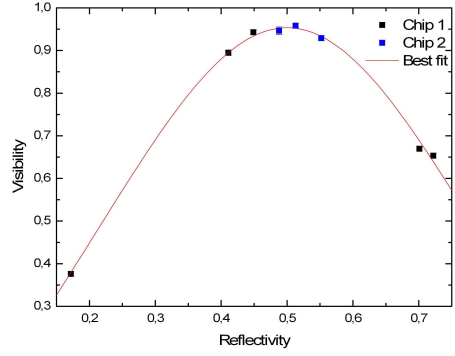}
\caption{Quantum interference visibility as a function of coupling ratio. Hong--Ou--Mandel interference visibility for couplers of various reflectivities $\eta$. Error bars are determined from fits such as those in Fig.~\ref{dip} and are comparable to the point size.}
\label{vis}
\end{center}
\end{figure}

\vspace{6pt}
We measured the performance of the DWQC devices using the 2-photon setup shown schematically in Fig.~\ref{schematic}(d). Figure \ref{dip} shows the raw data for a HOM dip in a coupler with $\eta=0.5128\pm 0.0007$ (maximum theoretical visibility $V_{ideal}=0.9987\pm 0.0001$). The measured visibility is $0.958\pm 0.005$. Figure \ref{vis} shows the measured visibility $V$ as a function of the equivalent reflectivity $\eta$ for eight couplers on the two chips. The curve is a fit of Eq.~(\ref{visibility}), modified to include a single parameter to account for mode mismatch \cite{ro-pra-72-032306}. The average relative visibility for these eight couplers is $\overline{V_{rel}}=0.952\pm 0.005$, demonstrating high performance across all couplers on both chips.
{Using single-photons, we measured the effective reflectivity of a Mach-Zehnder interferometer and found that the reflectivity of the device was $\eta_{MZ}=0.960\pm 0.001$. This indicates that the error in written phase shift} in the interferometer was very close to zero {(of the order 10~nm).}

\begin{figure}
\begin{center}
        \begin{minipage}[t]{0.02\textwidth}
            \vspace{0pt}
            \hspace{-12pt}
            (a)
        \end{minipage}
        \begin{minipage}[t]{0.40\textwidth}
            \vspace{0pt}
            \hspace{0pt}
            \includegraphics*[width=1.00\textwidth]{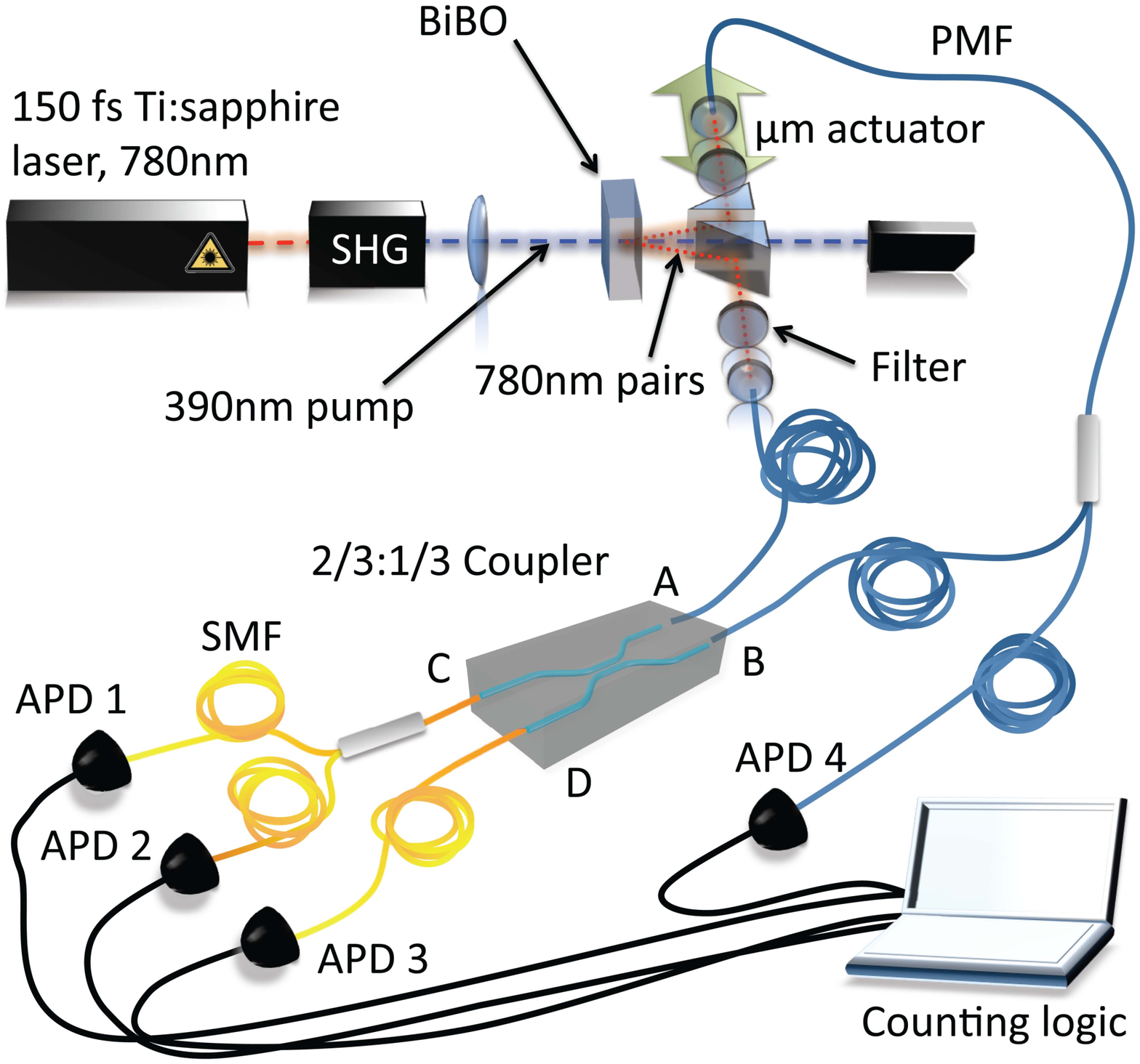}
        \end{minipage}
        \begin{minipage}[t]{0.02\textwidth}
            \vspace{0pt}
            \hspace{12pt}
            (b)
        \end{minipage}
        \begin{minipage}[t]{0.45\textwidth}
            \vspace{0pt}
            \hspace{0pt}
            \includegraphics*[width=1.00\textwidth]{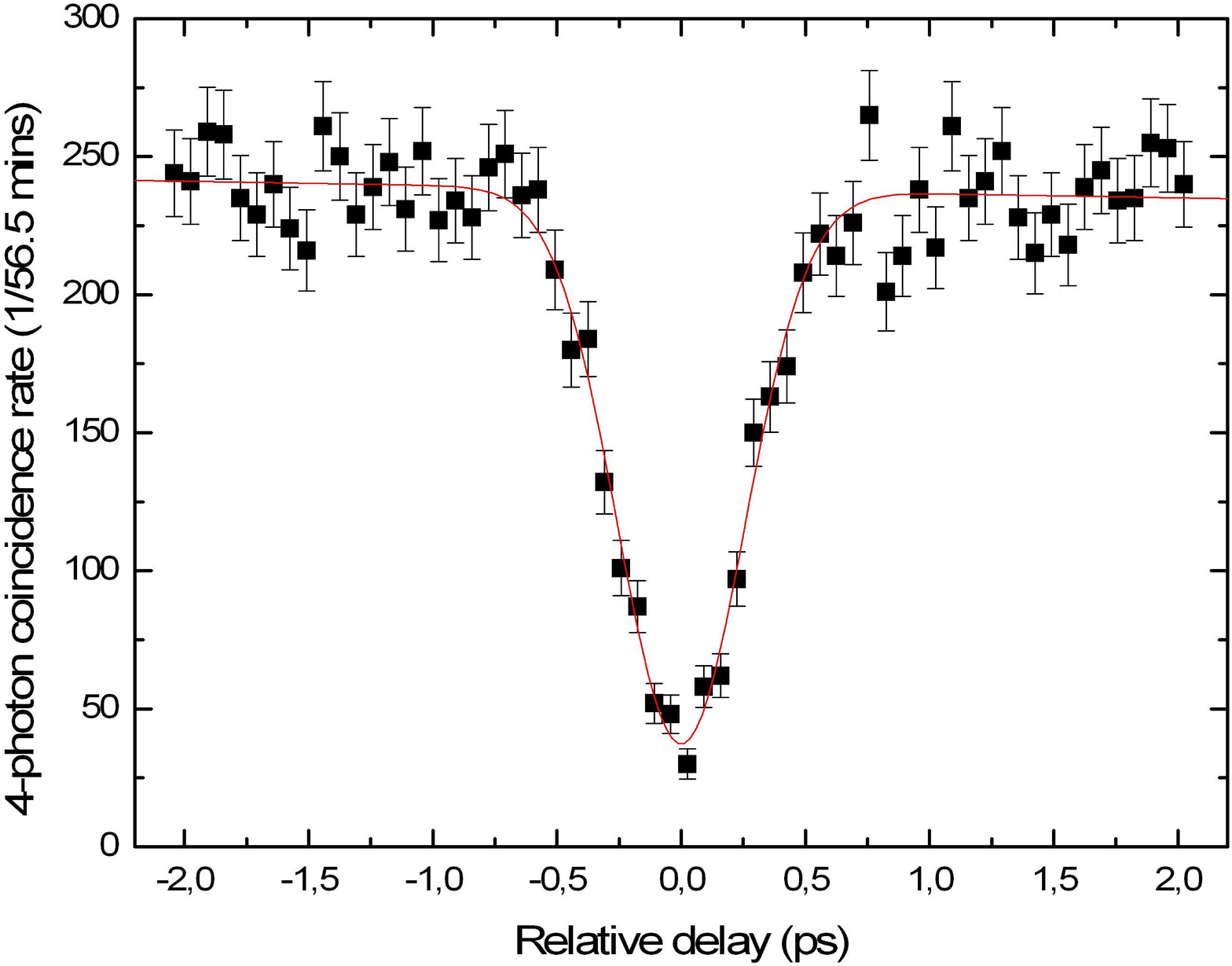}
        \end{minipage}
\caption{Generalized quantum interference with three photons. (a) Measurement setup consisting of a SPDC source based on a frequency doubled Ti:Sapphire \textit{fs} oscillator. (b) Generalized HOM dip for three photons. The number of coincident detections are shown as a function of the arrival delay between the two interfering photons.}
\label{21dip}
\end{center}
\end{figure}

{Using the 4-photon source shown in Fig.~\ref{21dip}(a) we were able to generate the three-photon entangled state described in Eqn.~(\ref{21hom}).} Figure~\ref{21dip}(b) shows the generalized HOM dip observed in a $\eta=0.659$ reflectivity DWQC coupler. The visibility of this dip is  $V_{rel}=0.84\pm0.03$ {which surpasses the value of $V=0.78\pm0.05$ previously observed in a bulk optical implementation \cite{sa-prl-96-083601})}. We believe our visibility to be limited by a small amount of temporal distinguishability of the photons produced in the source, nominally in the $|2\rangle_A$ state, and not by the waveguide device. {These results demonstrate the enhanced stability afforded by guided-wave circuits and are the first demonstration of multi-photon interactions in an integrated optics platform.}
\vspace{6pt}
\section{Conclusions}
\vspace{6pt}
DWQCs overcome many of the limitations of standard lithographic approaches: while the devices described here were written in 2D, extension to a 3D architecture is straightforward; single devices can be made with short turnaround for rapid prototyping; and the direct-write technique easily produces devices with circular mode profiles, the size and elipticity of which can be adjusted by changing the laser focusing conditions. This ability to tailor the guided mode will enable production of waveguides that better match fibre modes and reduce photon losses, and could be combined with the ongoing development of waveguide \cite{fu-oe-15-12769,zh-oe-15-10288} or fibre \cite{fu-prl-99-120501} based photon sources. Fabrication of sophisticated integrated quantum photonic circuits can now be achieved with only an ultrafast laser system rather than state-of-the-art semiconductor processing facilities. {Quantum circuits for photons, regardless of their application, comprise multi-path and multi-photon interferometers involving generalized quantum interference at a beam splitter (or directional coupler). High visibility interference such as that demonstrated here is therefore crucial to realizing future quantum technologies and} the next generation of fundamental quantum optics experiments.

\vspace{12pt}
\hspace{-12pt}
\textbf{Acknowledgments}\\
This work was supported by the Australian Research Council through their centres of excellence program, the UK EPSRC, QIP IRC, the {Macquarie University Research Innovation Fund} and the US IARPA.
\end{document}